\documentstyle[multicol,prd,aps]{revtex}

\begin{document}
\draft
\title{   Radiation reaction and the self-force
          for a point mass in general relativity}
\author{  Steven Detweiler }
\address{ Department of Physics, PO Box 118440, University of Florida,
          Gainesville, FL 32611-8440}
\maketitle

\begin{abstract}

A point particle of mass $\mu$ moving on a geodesic creates a
perturbation $h^\mu$, of the spacetime metric $g^0$, that
diverges at the particle. Simple expressions are given for the
singular $\mu/r$ part of $h^\mu$ and its quadrupole distortion
caused by the spacetime. Subtracting these from $h^\mu$ leaves a
remainder $h^R$ that is $C^1$. The self-force on the particle
from its own gravitational field corrects the worldline at
$O(\mu)$ to be a geodesic of $g^0+h^R$. For the case that the
particle is a small non-rotating black hole, an approximate
solution to the Einstein equations is given with error of
$O(\mu^2)$ as $\mu\rightarrow0$.

\end{abstract}

\begin{multicols}{2}

Perturbation analysis provides one approach to understanding the
self-force and radiation reaction in general relativity.  This
begins with a background spacetime metric $g^0$ which is a
vacuum solution of the Einstein equations $G_{ab}(g^0)=0$. An
object of small mass $\mu$ then disturbs the geometry by an
amount $h^\mu = O(\mu)$ which is determined by the perturbed
Einstein equations with the stress-energy tensor of the object
being the source,
\begin{equation}
  E_{ab}(h^\mu) = - 8\pi T_{ab} +O(\mu^2),\;\mu\rightarrow0.
\label{EabTab}
\end{equation}
The superscript ${}^\mu$ is a reminder that $h^\mu$ is linear in $\mu$.
The linear
differential operator $E_{ab}$ is defined by
\begin{equation}
  E_{ab}(h) = - \frac{\delta G_{ab}}{\delta g_{cd}} h_{cd},
\label{Eabdef}
\end{equation}
which evaluates to \cite{MTW}
\begin{eqnarray}
  2E_{ab}(h) &=& \nabla^2 h_{ab} + \nabla_a \nabla_b h
           - 2 \nabla_{(a}\nabla^c h_{b)c}
\nonumber\\ & &
           + 2{R_a}^c{}_b{}^d h_{cd}
           + g_{ab} ( \nabla^c\nabla^d h_{cd} - \nabla^2 h ).
\label{Eab}
\end{eqnarray}
With a solution of {Eq.\ (\ref{EabTab})} it follows that
\begin{equation}
  G_{ab}(g^0+h^\mu) = 8 \pi T_{ab} + O(\mu^2).
\end{equation}
An integrability condition for {Eq.\ (\ref{EabTab})} results from the
Bianchi identity for $g^0+h^\mu$ and requires that $T$ be conserved in
the background geometry up to $O(\mu^2)$.

Formally, perturbation analysis at the second order is no more
difficult than at the first. But the integrability condition
for the second order equations is that $T$ be conserved not in
the background geometry, but in the first order perturbed
geometry.  Thus, before solving the second order equations, it
is necessary to change the stress-energy tensor in a way which
is dependent upon the first order metric perturbations. This
correction to $T$ is said to result from the ``self-force'' on
the particle from its own gravitational field and includes the
dissipative effects of what is often referred to as ``radiation
reaction'' as well as other nonlinear aspects of general
relativity.

To focus on those details of the self-force which are
independent of the object's structure we restrict the object to
be a point particle with no spin angular momentum or other
internal structure. The integrability condition at the first
order then implies that the worldline $\Gamma$ of the particle
is nearly a geodesic of $g^0$, with an acceleration of only
$O(\mu)$. But, the integrability condition at the second order
presents a difficulty. The particle is to move along a geodesic
of $g^0+h^\mu$, but $h^\mu$ scales as $\mu/r$ near the particle
and is not differentiable on $\Gamma$.

Mino, et al.\cite{Mino97} and Quinn and Wald \cite{QuinnWald97} resolve
this difficulty  with a Green's function approach to {Eq.\
(\ref{EabTab})}. The formal Hadamard expansion of the Green's function
near the worldline of the particle identifies the ``instantaneous'' and
``tail'' parts of $h^\mu$. And, Mino et al. use a matched asymptotic
expansion to show that the particle moves along a geodesic of
$g^0+h^\mu_{\text{tail}}$; by construction $h^\mu_{\text{tail}}$ is
differentiable on the worldline as required by the geodesic equation.
However, their analysis provides no simple method for determining this
tail part.

Alternatively, we resolve the difficulty by finding the source part of
the metric perturbation $h^S$, which consists of the singular $\mu/r$
part plus its quadrupole distortion caused by the background geometry.
{Eqs.\ (\ref{hS})}-(\ref{h2mu}) give a simple expression for $h^S$. Then
we show that the remainder, $h^R \equiv h^\mu - h^S$, is $C^1$ and, using
matched asymptotic expansions, that the $O(\mu)$ effect of the self-force
adjusts the worldline of the particle to be a geodesic $\Gamma^\prime$ of
$g^0+h^R$. The consistency of our matched asymptotic expansions with
those of Ref.\ \cite{Mino97} imply that $h^R$ must be equivalent to the
``tail'' part of the metric perturbation from the Green's function, up to
a gauge transformation and terms of $O(\mu r^2)$, which do not effect the
$O(\mu)$ correction to the worldline.

The source field $h^S$ is best described with coordinates in which the
background geometry looks as flat as possible near the geodesic $\Gamma$.
A {\it normal} coordinate system, $x^a = (t,x,y,z)$, can be found
\cite{MTW} where, on $\Gamma$, the metric and its first derivatives match
the Minkowski metric, and the coordinate $t$ measures the proper time.
Normal coordinates for a geodesic are not unique, and we use particular
coordinates introduced by Thorne and Hartle \cite{ThorneHartle85} in
their discussion of external multipole moments of a vacuum solution of
the Einstein equations where
\begin{equation}
  g^0_{ab} = \eta_{ab} + {}_2H_{ab} + O(r^3/{\cal R}^3),
  \; r/{\cal R}\rightarrow0,
\label{H's}
\end{equation}
with
\begin{eqnarray}
  {}_2H_{ab}dx^a dx^b & = &
         - {\cal E}_{ij} x^i x^j ( dt^2 + \delta_{kl} dx^k dx^l )
\nonumber\\ &&
         {} + \frac{4}{3} \epsilon_{kpq}{\cal B}^q{}_i x^p x^i dt\, dx^k.
\label{H2}
\end{eqnarray}
And, ${\cal E}$ and ${\cal B}$ are spatial, symmetric, tracefree and
related to the Riemann tensor evaluated on $\Gamma$ by ${\cal E}_{ij} =
R_{titj}$ and ${\cal B}_{ij} = \epsilon_i{}^{pq}R_{pqjt}/2$;
 and, ${\cal R}$ is a representative length scale of the background
geometry---the smallest of the radius of curvature, the scale of
inhomogeneities, and the time scale for changes in curvature along
$\Gamma$, then ${\cal E}_{ij}$ and ${{\cal B}^q}_i$ are $O(1/{\cal R}^2)$;
 also $(r, \theta, \phi)$ are defined in the usual way in terms of
$(x, y, z)$; the indices $i$, $j$, $k, \ldots$ are spatial and raised and
lowered with $\delta_{ij}$.

If a small non-rotating black hole moves along $\Gamma$, then its geometry
is perturbed by tidal forces,
\begin{equation}
  g_{ab}^{\text{pert}} = g_{ab}^{\text{Schw}} + {}_2h_{ab}
\end{equation}
through terms of $O(r^2/{\cal R}^2)$, where ${}_2h$ is a solution of
\begin{equation}
E^{\text{Schw}}_{ab}({}_2h) = 0 \label{hlab}
\end{equation}
with the boundary conditions that the perturbation be well behaved on the
event horizon and that ${}_2h \rightarrow {}_2H$ in the {\it buffer
region} \cite{ThorneHartle85}, where $\mu \ll r \ll {\cal R}$. Both
${}_2H$ and ${}_2h$ consist of $\ell=2$ tensor harmonics in the
Regge-Wheeler gauge \cite{RWVZ}; the angular dependence is through
$x^ix^j{\cal E}_{ij}$ and $\epsilon_{kpq}{\cal B}^q{}_i x^p x^i$. For $r
\ll {\cal R}$, ${}_2h$ is governed by a wave equation with a potential
barrier. In the time independent limit this admits an analytic solution
\cite{RWVZ}
\begin{eqnarray}
 {}_2h_{ab}dx^a dx^b & = &
         - {\cal E}_{ij}x^i x^j \big[ (1-2\mu/r)^2 dt^2 + dr^2
\nonumber  \\ & &
         {} + (r^2-2\mu^2)(d\theta^2 + \sin^2\theta d\phi^2) \big]
\nonumber  \\ & &
         {} + \frac{4}{3} \epsilon_{kpq}{\cal B}^q{}_i x^p x^i (1-2\mu/r) dt\,
         dx^k
\label{h2ab}
\end{eqnarray}
which is well behaved on the event horizon and matches ${}_2H$ when
$\mu\ll r$. See Ref.\ \cite{MDA} for the case of a small black hole in
the vicinity of a larger black hole.

Time dependence of ${}_2H$ induces a quadrupole moment on the black hole,
but the resulting acceleration of the world line is smaller than
$O(\mu/{\cal R}^2)$. Generally, the timescale of ${}_2H$ is ${\cal R}$ and
corresponds to a low frequency for the black hole, $\omega\mu =
O(\mu/{\cal R}) \ll 1$. And two independent solutions for the metric
perturbation ${}_2h$ are standing waves very near $r=2\mu$ but behave as
$r^{2}$ and $1/r^3$, for $\mu \ll r \ll {\cal R}$. The proper solution is
a traveling wave into the hole created by a linear combination of these
two independent solutions having comparable magnitudes when $r \approx
2\mu$. Thus, for $r \gg \mu$ this linear combination is approximately
given by {Eq.\ (\ref{h2ab})}, which scales as $r^2$, plus a $1/r^3$
contribution from the induced quadrupole moment, ${\cal I}^{ab} =
O(\mu^5/{\cal R}^2)$, stemming from the time dependence. This
contribution to the quadrupole field couples to the background octupole
field and accelerates the worldline\cite{Zhang85} by $\sim{{\cal
E}^i}_{ab}{\cal I}^{ab}/\mu = O(\mu^4/{\cal R}^5)$ which is too small to
be important in this analysis.

In the buffer region, where $\mu \ll r \ll {\cal R}$, the geometry of a
point particle moving through the background should be equally well
described either by the background metric perturbed by $\mu$, or by the
leading $\mu/r$ terms of the Schwarzschild metric perturbed by weak tidal
forces.  In this region, then, the background metric perturbation $h^\mu$
is approximately the part of $g^{\text{pert}}$ which is linear in $\mu$;
and this part is the source field,
\begin{equation}
  h^S \equiv {}_0h^\mu + {}_2h^\mu,
\label{hS}
\end{equation}
where
\begin{equation}
  {}_0h^\mu_{ab}dx^a dx^b = 2\mu/r(dt^2 + dr^2)
\label{h0}
\end{equation}
is the $\mu/r$ part of the Schwarzschild metric, and
\begin{equation}
  {}_2h^\mu_{ab}dx^adx^b = \frac{4\mu}{r}{\cal E}_{ij}x^i x^j dt^2
      - \frac{8\mu}{3r} \epsilon_{kpq}{\cal B}^q{}_{i}x^p x^i dt\, dx^k
\label{h2mu}
\end{equation}
is the $\mu r/{\cal R}^2$ part of ${}_2h$ from {Eq.\
(\ref{h2ab})}. A split of $h^\mu$, from {Eq.\ (\ref{EabTab})},
into two pieces,
\begin{equation}
  h^\mu = h^S + h^R,
\end{equation}
reveals just how accurately $h^S$ approximates $h^\mu$ by
consideration of the remainder $h^R$. From {Eq.\ (\ref{EabTab})}
\begin{equation}
  E_{ab}(h^R) =  - E_{ab}(h^S) - 8\pi T_{ab}.
\label{EabhR}
\end{equation}
And, direct evaluation shows that
\begin{equation}
E_{ab}(h^S)+8\pi T_{ab} = O(\mu /{\cal R}^3), \; r\rightarrow0,
\label{EabhS}
\end{equation}
for a point particle stress tensor, and the source of {Eq.\
(\ref{EabhR})} is finite but not continuous at $r=0$.
 This last result is understandable---in an expansion
of $E_{ab}(h^S)$ in powers of $\mu$ and $1/{\cal R}$, all of the
$\mu/{\cal R}^2$ terms would also appear in a similar expansion of
$E^{\text{Schw}}_{ab}({}_2h)$. And this latter expansion is zero from the
definition of ${}_2h$. Thus, the source of {Eq.\ (\ref{EabhR})} is
$O(\mu/{\cal R}^3)$ as $r\rightarrow0$.

The solution of {Eq.\ (\ref{EabhR})} for $h^R$ with reasonable boundary
conditions is $C^1$.  If the metric is sufficiently differentiable and
there are no unreasonable boundary conditions at large distances, then
$h^R$ is differentiable away from the geodesic $\Gamma$. For if it were
not, then the discontinuities in the derivatives would propagate along the
characteristics of the hyperbolic operator $E_{ab}$ and would have
originated either on $\Gamma$ or on some boundary; we consider such
discontinuities emanating from a boundary to be unreasonable boundary
conditions. And, in a neighborhood of $\Gamma$ the geometry can be
smoothly mapped to flat spacetime with the operator $E_{ab}$ (in the
Lorentz gauge) being smoothly mapped to the flat spacetime wave operator
which, when integrated twice, smoothes a slowly changing and finite but
discontinuous (on $\Gamma$) source into a $C^1$ solution.
Thus the difference between $h^\mu$ and $h^S$ is a $C^1$
tensor field $h^R$.

The split of $h^\mu$ into $h^S$ and $h^R$ contains some arbitrariness.
Any piece of $O(\mu r^2/{\cal R}^3)$ or with higher powers of $r/{\cal
R}$ can be moved between $h^S$ and $h^R$ without affecting either {Eq.\
(\ref{EabhR})}, the differentiability of $h^R$, or the $O(\mu)$ effect of
the self-force which changes the worldline to be a geodesic
$\Gamma^\prime$ of $g^0+h^R$. Furthermore, a gauge transformation $y^a =
x^a + \xi^a$ for any $\xi^a$ that is $O(r^3/{\cal R}^2)$ gives a new
normal coordinate system; and we state without details that the
corresponding change in $h^S$ is only $O(\mu r^2/{\cal R}^3)$ with
vanishing derivatives on the worldline. Thus $\Gamma^\prime$ is
independent of the normal coordinate system in use.

But, it is not sufficient to have $h^S$ just consist of ${}_0h^\mu$. If
${}_2h^\mu$ were not included, then $E_{ab}(h^S)+8\pi T_{ab}$ would be
singular $\sim \mu/r{\cal R}^2$ as $r\rightarrow0$. The resulting $h^R$
would not be differentiable on the worldline, and some version of
averaging around the particle would be required to make sense of the
effects of the self-force. Thus, it is necessary to include ${}_2h^\mu$
as part of $h^S$.

Above, we mentioned that the worldline of a small particle through the
background is a geodesic of $g^0+h^R$ when the $O(\mu)$ corrections
are included.  We now justify this statement by replacing the particle
with a small, non-rotating black hole and considering a sequence of
metrics $g(\mu)$ which are solutions of the vacuum Einstein equations,
and parameterized by $\mu$ with $g(0) = g^0$. Our focus is on the
$O(\mu)$ behaviors of $g(\mu)$ and the worldline in the limits
$\mu\rightarrow0$ and the worldline approaching $\Gamma$.
 Matching asymptotic expansions in the buffer
region provides both the correction to the worldline as well as an
approximation to $g(\mu)$, {Eq.\ (\ref{gmatch})} below, which is uniformly
valid with an error of $O(\mu^2/{\cal R}^2)$, as $\mu\rightarrow0$.

This matching relies upon a choice of coordinates similar to that of
{Eq.\ (\ref{H's})} but for the metric $g^0+h^R$ being expanded about its
geodesic $\Gamma^\prime$; these ``primed'' normal coordinates differ from
the original by $O(\mu)$.  In this discussion of matching, $r$ is now
associated with $\Gamma^\prime$ and the primed coordinates.

In the buffer region $g(\mu)$ is best described in a fashion introduced
by Thorne and Hartle \cite{ThorneHartle85} as a sum of terms of positive
powers of the small numbers $\mu/r$ and $r/{\cal R}$,
\begin{equation}
\begin{array}{cccccccccccccc}
  g(\mu) & \sim  & \eta^\prime & \& & 0   & \& & {}_2H^\prime
                         & \& & \cdots
\\ & \& & \mu/r      & \& & \mu /{\cal R}      & \& & \mu r /{\cal R}^2
                     & \& & \cdots
\\ & \& & \mu^2/r^2  & \& & \mu^2 /r{\cal R}     & \& & \mu^2  /{\cal R}^2
                     & \& & \cdots
\\ &     \& & \cdots,
\end{array}
\label{tableau}
\end{equation}
where $\&$ means ``and a term of the form $\ldots$'' In this tableau, part
of any term can be moved diagonally, up and to the left, to be absorbed
into a dominating term with the same $r$ behavior.

Just outside the buffer region, where $\mu \ll r < {\cal R}$, $g(\mu)$ is
approximately the background geometry perturbed by $\mu$. The top row of
the tableau consists of the expansion of $g^0$ about $\Gamma^\prime$ in
powers of $r/{\cal R}$ along with some parts of a similar expansion of
$h^R$. The $\mu/{\cal R}$ and $\mu r /{\cal R}^2$ terms from $h^R$ are
absorbed into the first two elements of the top row, which are then
displayed as $\eta^\prime$ and zero because the primed coordinates are
normal. The $r^2/{\cal R}^2$ term of the tableau is set to ${}_2H^\prime$,
a tensor whose components in the primed coordinates are the same as those
of ${}_2H$ in the original coordinates. The difference between
${}_2H^\prime$ and ${}_2H$ is of $O(\mu r^2/{\cal R}^3)$ and is
subtracted from the fourth term in the second row of the tableau. The
rest of the first row consists precisely of the corresponding terms in
the expansion of $g^0$. And the remainder of the expansion of $h^R$, from
the $\mu r^2/{\cal R}^3$ term outward, contribute to the corresponding
terms of the second row. Thus the entire first row along with the fourth
and greater terms of the second row sums to $g^0+h^R$.

Just inside the buffer region, where $2\mu < r\ll{\cal R}$, $g(\mu)$ is
approximately the Schwarzschild geometry perturbed by background tidal
forces. The first column of the tableau, containing no ${\cal R}$, is an
expansion of the Schwarzschild geometry in powers of $\mu/r$. The second
column, linear in $1/{\cal R}$, sums to a dipole perturbation of the
Schwarzschild geometry. But, the top element of the second column is
zero, so all elements of the second column are zero. The top term in the
third column, ${}_2H^\prime$, when added to the rest of the third column
gives ${}_2h^\prime$, the quadrupole perturbation of the black hole
caused by tidal forces. Thus, the first three elements of the top row
determine the entire first three columns of this tableau by the
expansions of $g^{\text{Schw}}$ and ${}_2h^\prime$ in powers of $\mu/r$.

Now, $g^0+h^R$ is an accurate approximation of $g(\mu)$ when $\mu
\ll r$, and $g^{\text{Schw}}+{}_2h^\prime$ is an accurate
approximation when $r \ll {\cal R}$ and with $g^{\text{Schw}}$ centered on
$\Gamma^\prime$. These approximations overlap in the buffer region where
\begin{equation}
  g^0_{ab}+h_{ab}^R = \eta_{ab}^\prime + {}_2H_{ab}^\prime
            + O(r^3/{\cal R}^3),
\label{g0}
\end{equation}
and
\begin{equation}
  g_{ab}^{\text{Schw}}+ {}_2h_{ab}^\prime
                   = \eta_{ab}^\prime + {}_2H_{ab}^\prime + O(\mu/r)
\label{gpert}
\end{equation}
match asymptotically. In the restricted region $\mu/r \ll r^2/{\cal R}^2
\ll 1$, the displayed term ${}_2H^\prime=O(r^2/{\cal R}^2)$ is small yet
much larger than either of the remainder terms, $O(r^3/{\cal R}^3)$ and
$O(\mu/r)$, as $\mu/{\cal R}\rightarrow0$. This is the hallmark of matched
asymptotic expansions.  If the worldline of the black hole were not a
geodesic of $g^0+h^R$ then {Eq.\ (\ref{g0})} would necessarily contain a
term of $O(r/{\cal R})$. But an explicit $O(r/{\cal R})$ term in {Eq.\
(\ref{gpert})} would be the dominant term of a dipole perturbation, and
such a dipole vacuum perturbation of the Schwarzschild geometry is always
removable by a gauge transformation \cite{RWVZ}. Thus, this asymptotic
matching is only successful when the worldline $\Gamma^\prime$ is a
geodesic of $g^0+h^R$ up to an acceleration of $O(\mu^2/{\cal R}^3)$ in
the limit that $\mu/{\cal R} \rightarrow 0$.

A concise description of this matched geometry is
\begin{eqnarray}
  g_{ab}(\mu) &=&(g^0_{ab}+h_{ab}^R)
    + (g_{ab}^{\text{Schw}} + {}_2h_{ab}^\prime)
\nonumber \\
    &&{}- (\eta_{ab}^\prime + {}_2H_{ab}^\prime) + O(\mu^2/{\cal R}^2),
                    \; \mu/{\cal R}\rightarrow0.
\label{gmatch}
\end{eqnarray}
For $r\ll{\cal R}$, the first and third terms on the right nearly cancel and
give $g(\mu)\approx g^{\text{Schw}} + {}_2h^\prime$, the first
three columns of the tableau. For $\mu \ll r$ the second and third terms
on the right yield $h^{S\prime}+O(\mu^2/r^2)$. And $g(\mu) \approx
g^0+h^R+h^{S\prime}$, the top two rows of the tableau. The error in
this approximation comes from the $\mu^2r/{\cal R}^3$ term in the tableau,
the dominant term not fixed in the above discussion.  This term and the
rest of the third row sum to a contribution of $O(\mu^2/{\cal R}^2)$. Thus,
the matched asymptotic expansions give an approximation to $g(\mu)$ in
{Eq.\ (\ref{gmatch})} which is uniformly valid
with error of $O(\mu^2/{\cal R}^2)$ as $\mu\rightarrow0$.

An application of this approach, in conjunction with Fourier-harmonic
decomposition, determines the $O(\mu)$ corrections to geodesic motion for
a small non-rotating mass in orbit about a much larger non-rotating black
hole. First, {Eq.\ (\ref{EabTab})} is solved for $h^\mu$ using the usual
metric perturbation analysis of the Regge-Wheeler formalism \cite{RWVZ}.
This involves decomposing $T$ into its Fourier-harmonic modes; then, the
inhomogeneous Regge-Wheeler or Zerilli equation\cite{RWVZ} is integrated
numerically to determine the radial dependence of the modes of $h^\mu$.
And, given the appropriate coordinate transformation, the components of
$h^S$ can be transformed from {Eqs.\ (\ref{h0})} and (\ref{h2mu}) to the
usual Schwarzschild coordinates and then numerically decomposed into their
Fourier-harmonic modes. Now, $h^R$ and its derivatives can be constructed
as the sum over modes of the difference between $h^\mu$ and $h^S$. And,
the $O(\mu)$ effect on the worldline of the small mass can be calculated
as a change from a geodesic of $g^0$ to a geodesic of $g^0+h^R$.

This application depends upon the transformation between the normal
coordinates of {Eq.\ (\ref{H's})} and Schwarzschild coordinates. Manasse
and Misner \cite{ManasseMisner} give a prescription for finding
Fermi-normal coordinates for any geodesic; and Zhang \cite{Zhang86} gives
a gauge transformation from these to the coordinates of {Eq.\
(\ref{H's})}. But {Eq.\ (\ref{H's})} only determines the normal
coordinates near the worldline and up to terms of $O(r^4/{\cal R}^3)$.
So, the normal coordinates must be extended to cover the Schwarzschild
manifold for all radii in the vicinity of the orbit to make the mode
decomposition of $h^S$ possible. Fortunately, the details of this
extension are not important because $h^R$ is known to be differentiable
and, therefore, easily describable in terms of a sum over its modes.
While the amplitude of each individual mode of $h^S$ does depend upon the
extension of the coordinates away from the worldline, the reconstruction
of $h^R$ near the worldline by the sum over modes is independent of this
extension.

This analysis shows that simply removing the divergent monopole part
${}_0h^\mu$ from the metric perturbation $h^\mu$ leaves a
nondifferentiable remainder. But, if the quadrupole distortion of the
monopole is also removed from $h^\mu$, then the remainder, $h^R = h^\mu -
{}_0h^\mu - {}_2h^\mu$, is differentiable and suitable for calculating
$O(\mu)$ effects on the worldline.

The octupole term ${}_3h^\mu$ can also be removed from $h^\mu$. Thorne
and Hartle \cite{ThorneHartle85} extend the coordinates in {Eq.\
(\ref{H's})} to include $O(r^3/{\cal R}^3)$ terms explicitly, their
 Eqs. (A1) and (A2). The
time independent solution of $E^{\text{Schw}}_{ab}({}_3h) = 0$ which is
well behaved on the event horizon and properly matches the
$O(r^3/{\cal R}^3)$ terms of Ref.\ \cite{ThorneHartle85} is
\begin{eqnarray}
 {}_3h_{ab}&& dx^a  dx^b  =
         - \frac{1}{3}{\cal E}_{ijk}x^i x^j x^k \big[
                \big(1-\frac{2\mu}{r}\big)^2\big(1-\frac{\mu}{r}\big) dt^2
\nonumber  \\ & &
         + \big(1-\frac{\mu}{r}\big)dr^2
         + \big(r^2-2\mu r + \frac{4\mu^3}{5r}\big)
                     (d\theta^2 + \sin^2\theta \, d\phi^2) \big]
\nonumber  \\ & &
         + \frac{2}{3} \epsilon_{kpq}{\cal B}^q{}_{ij} x^p x^i x^j
           \big(1-\frac{2\mu}{r})(1-\frac{4\mu}{3r}\big) dt\, dx^k.
\label{h3ab}
\end{eqnarray}
And, ${}_3H$ and ${}_3h^\mu$ are the $r^3/{\cal R}^3$ and $\mu r^2/{\cal
R}^3$ parts of ${}_3h$.  If ${}_3h^\mu$ is also removed from $h^\mu$ then
the remainder is $C^2$, the matching is extended through the $1/{\cal
R}^3$ terms, but the overall error of the matched geometry is still
$O(\mu^2/{\cal R}^2)$ and there is no effect on the worldline.

Interesting discussions with Bernard Whiting, Eric Poisson, and Yasushi
Mino greatly clarified many aspects of this letter. This work was
supported in part by NASA grant NAGW-4864 with the University of Florida.

\bibliographystyle{prsty}

\end{multicols}
\end{document}